\documentclass[aps,prl,superscriptaddress,showpacs,floatfix,twocolumn]{revtex4}

\usepackage{graphicx}
\usepackage{epstopdf}
\usepackage{dcolumn}
\usepackage{bm}
\usepackage{units}

\begin{document}


\title{Production of Mesons and Baryons at High                       
        Rapidity and High $P_{T}$ in Proton-Proton  Collisions at $\sqrt{s} = \unit[200]{GeV}$}

\newcommand{\bnl}           {$\rm^{1}$}
\newcommand{\ires}          {$\rm^{2}$}
\newcommand{\kraknuc}       {$\rm^{3}$}
\newcommand{\krakow}        {$\rm^{4}$}
\newcommand{\newyork}       {$\rm^{6}$}
\newcommand{\nbi}           {$\rm^{7}$}
\newcommand{\texas}         {$\rm^{8}$}
\newcommand{\bergen}        {$\rm^{9}$}
\newcommand{\bucharest}     {$\rm^{10}$}
\newcommand{\kansas}        {$\rm^{11}$}
\newcommand{\oslo}          {$\rm^{12}$}
\newcommand{\spaceSc}          {$\rm^{13}$}
\newcommand{\pt} {$p_{T}$}

\author{
  I.~Arsene\oslo, 
  I.~G.~Bearden\nbi, 
  D.~Beavis\bnl, 
  S.~Bekele\kansas, 
  C.~Besliu\bucharest, 
  B.~Budick\newyork, 
  H.~B{\o}ggild\nbi, 
  C.~Chasman\bnl, 
  C.~H.~Christensen\nbi, 
  H.H.~Dalsgaard\nbi,
  R.~Debbe\bnl, 
  J.~J.~Gaardh{\o}je\nbi, 
  K.~Hagel\texas, 
  H.~Ito\bnl, 
  A.~Jipa\bucharest, 
  E.~B.~Johnson\kansas, 
  C.~E.~J{\o}rgensen\nbi, 
  R.~Karabowicz\krakow, 
  N.~Katrynska \krakow, 
  E.~J.~Kim\kansas, 
  T.~M.~Larsen\oslo, 
  J.~H.~Lee\bnl, 
  S.~Lindal\oslo, 
  G.~L{\o}vh{\o}iden\oslo, 
  Z.~Majka\krakow, 
  M.~Murray\kansas, %
  J.~Natowitz\texas, 
  B.~S.~Nielsen\nbi, 
  C.~Nygaard\nbi,
  R.~P\l aneta\krakow, 
  F.~Rami\ires, 
  F.~Renault\nbi, 
  C.~Ristea\nbi, 
  O.~Ristea\bucharest, 
  D.~R{\"o}hrich\bergen, 
  B.~H.~Samset\oslo, 
  S.~J.~Sanders\kansas, 
  R.~A.~Scheetz\bnl, 
  P.~Staszel\krakow, 
  T.~S.~Tveter\oslo, 
  F.~Videb{\ae}k\bnl, 
  R.~Wada\texas, 
  Z.~Yin\bergen, 
  H.~Yang\bergen, and
  I.~S.~Zgura\spaceSc\\ 
  The BRAHMS Collaboration \\ [1ex]
  \bnl~Brookhaven National Laboratory, Upton, New York 11973 \\
  \ires~Institut Pluridisciplinaire Hubert Curien  and Universit{\'e} Louis
  Pasteur, Strasbourg, France\\
  \krakow~Smoluchkowski Inst. of Physics, Jagiellonian University, Krakow, Poland\\
  \newyork~New York University, New York 10003 \\
  \nbi~Niels Bohr Institute, Blegdamsvej 17, University of Copenhagen, Copenhagen 2100, Denmark\\
  \texas~Texas A$\&$M University, College Station, Texas 17843 \\
  \bergen~University of Bergen, Department of Physics and Technology, Bergen, Norway\\
  \bucharest~University of Bucharest, Romania\\
  \kansas~University of Kansas, Lawerence, Kansas 66049 \\
  \oslo~University of Oslo, Department of Physics, Oslo, Norway\\
  \spaceSc~Institute for Space Science, Bucharest, Romania\\
 }

\date{\today}

\begin{abstract}

We present particle spectra for charged hadrons $\pi^\pm, K^\pm, p$  and $\bar{p}$ from
pp collisions at $\sqrt{s}=200$ GeV measured for the first time at forward rapidities (2.95 and 3.3). The kinematics of these measurements are skewed in a way that probes the small momentum fraction in one of the protons and large fractions in the other.  Large proton to pion ratios are
observed at values of transverse momentum that extend up to  4 GeV/c, where
protons have momenta up to 35 GeV.
Next-to-leading order perturbative QCD calculations
describe the production of pions and kaons well at these rapidities,  but fail to account for the large proton yields and small $\bar{p}/p$ ratios.

\end{abstract}

\maketitle

Even though proton-proton  collisions are often used as  a simple hadronic reference system to  disentangle nuclear effects  in p-A or A-A systems, much of that  interaction remains unknown. 
Quantum Chromodynamics (QCD),
the accepted theory of hadronic systems, has had impressive success in describing many aspects of these collisions, specially
when the underlying partonic processes are hard and a perturbative approach is possible. 
Such an approach starts to be applicable at the CERN ISR energies ( $\sqrt{s} = 30-60\ $GeV), but only  near mid-rapidity  \cite{ISRrapidity, Soffer}.   
Measurements of jet and neutral pion production from $\bar{p}+p$ collisions at higher energies at the SppS \cite{UA2rapidity} and the TEVATRON  \cite{D0rapidity} 
show that those systems can be well described perturbatively in a wide range of rapidities. 
The RHIC program at BNL has brought the focus back to p+p collisions at $\sqrt{s}=\unit[200]{GeV}$ 
with an impressive array of probes that range from fully identified hadrons in a wide rapidity range, to direct photons, heavy quark production and jets.
Cross sections measured at RHIC have been compared to calculations based on perturbative QCD (pQCD) where the partonic cross sections are calculated up to  Next-To-Leading-Order level of accuracy (NLO) \cite{Vogelsang, deFlorian}. The  measured yields of neutral pions and direct photons at mid-rapidity \cite{PHENIXpi0, PHENIXdirectGam},  as well as the production of neutral pions at high rapidity  are well reproduced by the calculations \cite{STARppForward}. 

Collider measurements at high rapidities ($y_{beam} = 5.4$ at $\sqrt{s}=\unit[200]{GeV}$ ) offer a window to 
an interesting 
region of the proton wave function; 
the kinematics of the collision at the partonic level is skewed such that 
one of the protons is probed at small values of fractional longitudinal momentum {\it x} by valence quarks from the other proton that have  higher values of {\it x}. 
The nucleon internal structure has been well measured for a wide range of   {\it x} values, and there are indications that new regimes could be reached within the {\it small-x} component of the wave function  where gluons are the most abundant partons 
\cite{CGC}. High rapidity measurements in p+p and p+A collisions are thus a useful   
tool to extend the applicability of QCD.  
These data also serve as a reference to understand the evolution of the
gold wave function with rapidity as observed in d+Au collisions, and the formation of a dense and
opaque colored medium in heavy ion collisions \cite{RdA2003, RHICdA,  BRAHMSdA}.

This letter reports the first measurement of unbiased  differential cross sections  for  identified charged pions, kaons, protons and their anti-particles at high rapidity in p+p  collisions at $\sqrt{s}=\unit[200]{GeV}$. 
The comparison of these data to a QCD framework furthers the understanding of the hadron-hadron interaction.


\begin{figure}[!ht]
  \resizebox{0.9\columnwidth}{!}
  {\includegraphics{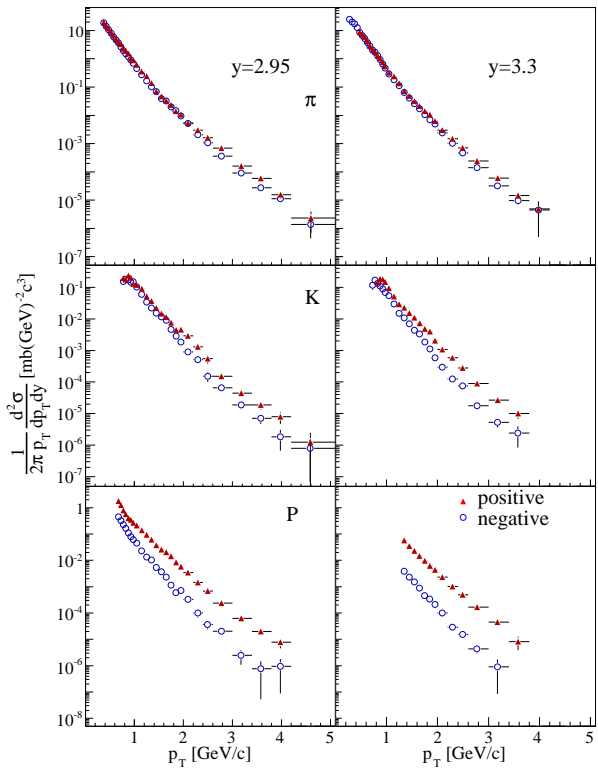}}
  \caption{\label{fig:spectra} Invariant cross section
    distributions for pion, kaons protons and anti-protons  produced in  p+p
    collisions at $\sqrt{s} = \unit[200]{GeV}$ at rapidities
    y=2.95 (left panels) and  y = 3.3 (right panels).  In all panels, positive charged particles are shown with  filled triangles (red in on-line version) and negative ones with open circles (blue in on-line version). The errors displayed 
in these plots are statistical. }
\end{figure}

The data were taken with the BRAHMS forward spectrometer set at
$4^{\circ}$ and $2.3^{\circ}$ with respect to the beam \cite{BRAHMSNIM}. These settings correspond to y=2.95 and 3.30 respectively.
At these rapidities the geometrical acceptance and the particle identification of the experiment allow us to reach high  $p_{T}$ values (4-5 GeV/c compared with the kinematical limits of 10.4 and 7.4  GeV/c, respectively),
which  is essential to test the ability of NLO pQCD to reproduce the measured yields. Measurements at two rapidities provide a check on rapidly falling cross sections.

The minimum bias trigger used to normalize these measurements was defined with a set of Cherenkov radiators (CC) placed symmetrically with respect to the nominal interaction point and covering  pseudo-rapidities that range in absolute value from 3.26 to 5.25.
This trigger required that at least one hit is detected in both sides of the array.
We used events generated with PYTHIA \cite{Pythia} and a GEANT simulation of the BRAHMS setup to estimate that this trigger can access $70 \pm 5\%$  of the total inelastic proton-proton cross section of 41 mb. 
The data used for this analysis were collected with a spectrometer trigger defined by three scintillator hodoscopes, located at 4.6, 8.8 and 18.9 meters 
 away from the nominal interaction point. 
GEANT simulations of the experiment with input events  generated by PYTHIA  have
 been used to estimate the bias introduced by the CC detectors in the spectrometer trigger. The deduced correction is equal to
$13\pm 7\%$,  approximately independent of $p_T$ and rapidity, and was applied to all spectra. 
The CC detectors have good timing resolution and were also used  to define the position of the interaction along the beam line with a resolution of approximately 2 cm.  
The present analysis was done with charged particles that originated from  collisions of polarized protons with interaction vertices in the range of  $\pm  40$ cm.  The transverse polarization of each beam bunch (arranged to point up or down in  an alternating way) is estimated to have a negligible contribution to the inclusive cross-sections presented here:  
even though the bunch intensities for opposite polarizations can differ by as much as $\sim 10\%$,  the asymmetries of produced particles were measured and have small values (~2-8\%) resulting in a net effect smaller than 1\%. Invariant cross sections were extracted 
in  narrow ($\Delta y = 0.1) $  rapidity bins centered at y=2.95 and y = 3.3, respectively. Narrow rapidity bins are required to reduce the effects of rapidly changing cross sections. Each distribution is obtained from the merging of up to five magnetic field settings. 
The data are corrected for the spectrometer geometrical acceptance, multiple scattering, weak decays and absorption in the material along the path of the detected particles. Tracking and matching efficiencies for each of the 5 tracking stations in the spectrometer were calculated by constructing full tracks with only 4 stations and evaluating the efficiency in the $5^{th}$ station. The overall efficiency is about 80-90\% and is included in the extraction of the cross sections. Particles are identified by our Ring Imaging Cherenkov detector (RICH) \cite{RICHNIM}.   The efficiency of this  detector has been studied with pions
identified with a scintillator time-of-flight counter in an overlapping momentum range and reaches an upper value of  97\%. The low momentum part of the proton spectra is measured using the RICH in veto mode.

\begin{figure}[!ht] 
\resizebox{0.9\columnwidth}{!}
          {\includegraphics{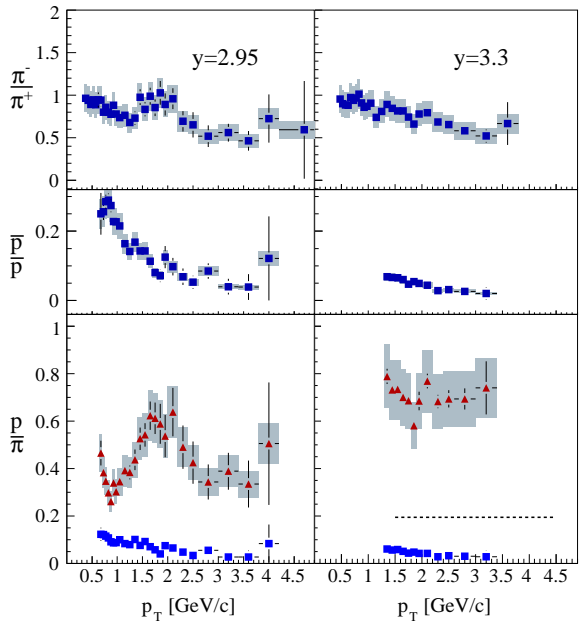}}
 \caption{\label{fig:ratio} Particle
ratios versus $p_T$ at y=2.95 and 3.3.
Top) $\pi^-/\pi^+$, Middle) ${\bar p}/p$  and Bottom) $p/\pi^+$ (red circles) and ${\bar p}/\pi^-$ (blue squares). The shaded rectangles indicate an overall systematic error (17\%) estimated for these ratios. The dashed line shows an upper limit for the
$(p+\bar{p})/(\pi^{+} + \pi^{-})$ ratio from $e^+ e^-$ collisions.}
         
\end{figure}

The momentum resolution of the spectrometers, $\delta p$, is calculated from the
angular resolution of the tracking detectors. With p  in GeV/c,
$\delta p/p = 0.0008p$ at the highest field setting.
From a comparison of the 3 independent spectrometer measurements and the momentum extracted from the ring radius in the RICH we estimate that our systematic error on the absolute momentum scale is less than 1\%. We have identified several sources of possible systematic errors and they are listed in Table I .

\begin{figure*}[!ht] 
  \resizebox{0.99\textwidth}{!}
          {\includegraphics{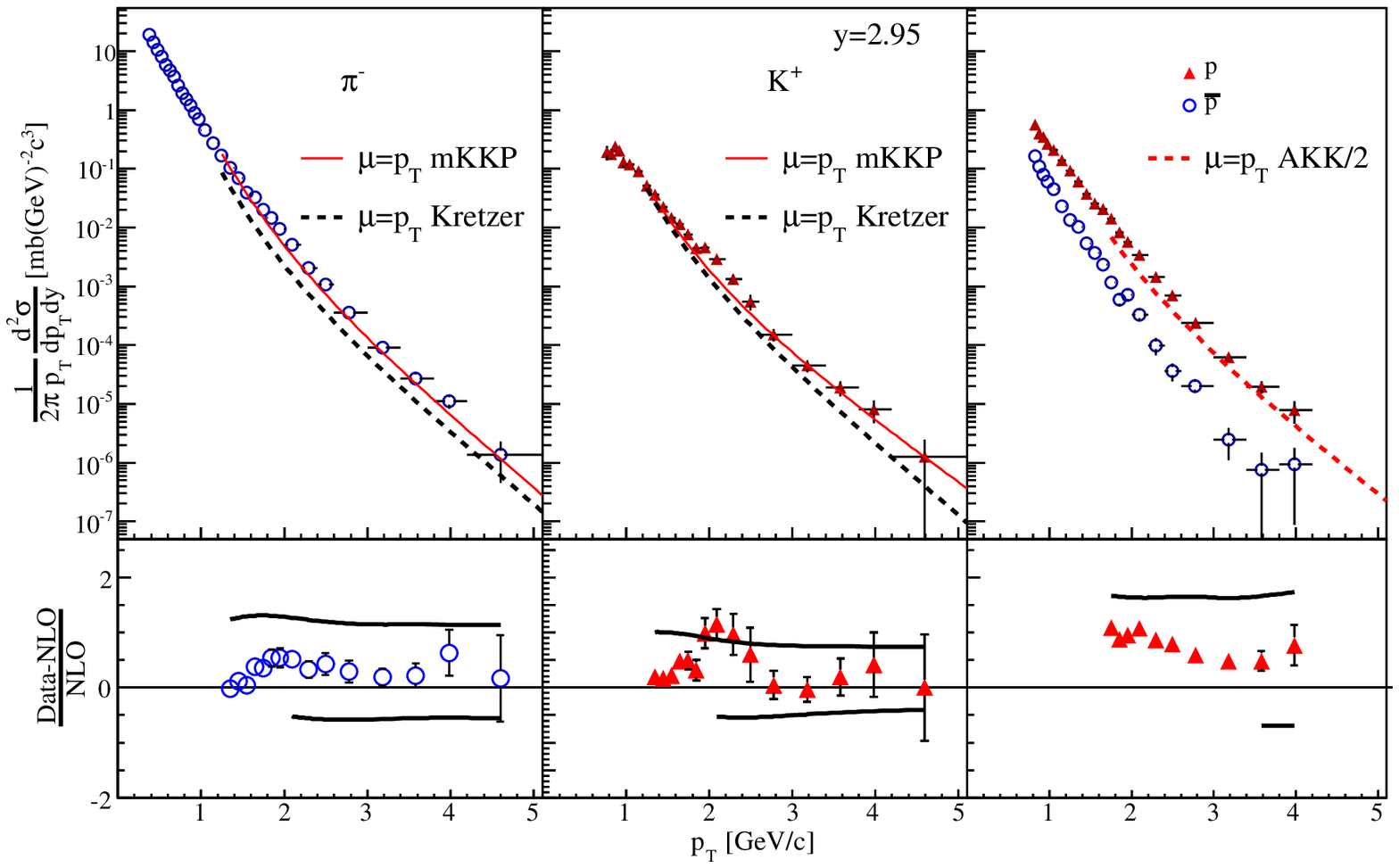}}
         \caption{\label{fig:NLO} Top) Comparison of invariant cross sections for $\pi^-$, $K^+$,
${\bar p}$ and $p$ at y=2.95 and  NLO calculations with factorization and renormalization  scales set equal to \pt.
The mKKP set of fragmentation functions (solid red line on-line) produce the 
best agreement with the $\pi^-$ and  $K^+$ data. The $p$ and  ${\bar p}$ are compared with the calculation using the AKK set divided by 2 
(dashed red line in on-line version), see text for details.
Bottom) Relative differences between data and calculations.
The top smooth curves show the effect of setting $\mu  = 2  p_{T}$ and the bottom curves $\mu$  = 1/2\pt .
For the baryons the (red) filled triangles show $p$ data vs the AKK/2 set.}
 
\end{figure*}


\begin{table*}
\begin {ruledtabular} 
\begin{tabular} {lccccccc}

$p_{T} [GeV/c]$ & vertex location   & normalization & PID ($\pi$) & PID(K) & PID(p) & decay and absorpt.  & p resol.\\
 \hline
0 - 1   & 5.6   & 7 & 5  & 5 & 8 & 2 & 1 \\
1 - 3   & 11.   & 7 & 8  & 8 & 8 & 2 & 1 \\
3 - 6   & 11    & 7 & 13 &18 & 13& 2 & 3 \\


\end{tabular}
\end{ruledtabular}
\caption{Estimated systematic errors shown as a per-cent of the measured cross-sections presented in this letter.} 
\label{tab:table1}
\end{table*}

Figure 1 shows the invariant cross sections for $\pi^\pm, K^\pm, p$ and
${\bar p}$ versus $p_T$ at y=2.95 and 3.3.
The data were placed at the average $p_T$ value calculated within the bin.
The upper panels of Fig. \ref{fig:ratio} show the $\pi^{-} / \pi^{+}$ ratio at y=2.95 and 3.3 respectively. Within systematic errors both ratios display a falling trend as a function of $p_{T}$. This may be an indication of the dominance of valence quark fragmentation at these rapidities and reflects the ratio of d to u quarks in the proton.
The $\bar{p} / p$ ratios shown in the middle panels are much smaller than unity. This is a clear indication that proton and anti-protons at these high rapidities are not produced by a common mechanism like $g \rightarrow p$ fragmentation. 
The bottom panels of Fig. \ref{fig:ratio} show the $p / \pi^{+}$ ratio as solid triangles and the $\bar{p} / \pi^{-}$ ratio with solid squares at y=2.95 and y=3.3. 
The dashed line is the maximum value for the measured  $(p+\bar{p})/(\pi^{+}+\pi^{-})$ ratio in  $e^{+} + e^{-}$ annihilation at 91.2 GeV \cite{epluseminusParam}.  
The difference between $p / \pi^{+}$ and $\bar{p} / \pi^{-} $ at both rapidities is remarkable and caused by an unexpectedly large proton yield at these rapidities. Such large yield may be
related to the mechanism that transfers a conserved quantity, the baryon number, from beam to intermediate rapidities i.e. baryon number transport. What remains as an open question is how  a mechanism that is thought to be mainly restricted to the longitudinal component of the momentum, gives these protons such high $p_{T}$. 

We have compared our measured differential cross-sections with NLO pQCD calculations \cite{Werner} evaluated at equal factorization and renormalization scales, $\mu \equiv \mu_{F}=\mu_{R}=p_{T}$. These calculations use the CTEQ6 parton distribution functions \cite{CTEQ6} and a modified version of the ``Kniehl-Kramer-P\"{o}tter" (KKP) set of fragmentation functions (FFs) \cite{KKP} referred to here as mKKP, as well as the ``Kretzer" (K) set \cite{Kretzer}.  
The KKP set includes functions that fragment into  the sums $\pi^{+}+\pi^{-}$, $K^{+}+K^{-}$ and $p+\bar{p}$.  
Modifications were necessary  to obtain functions producing the separate charges for both $\pi $ and $K$. 
These modifications are based in functional forms supported by measurement \cite{Reya}, and involve the following operations: to obtain the FFs producing $\pi^{+}$, the functions fragmenting favored light quarks  $u, \bar{d}$ into $\pi^{0}$ were multiplied by  $(1+z)$ ( e.g. $D_{u}^{\pi^{+}}=(
1+z)D_{u}^{\pi^{0}}$ with $D_{u}^{\pi^{0}}=\frac{1}{2}D_{u}^{\pi^{+}+\pi^{-}}$) where $z$ is the fraction of the parton momentum carried by the hadron, and $D_{u}^{\pi^{+}}$ the function fragmenting u quarks into positive pions. The functions fragmenting  unfavored quarks $\bar{u}, d$ into  $\pi^{0}$ are multiplied by $1-z$.  The same operation is done for  $\pi^{-}$, but this time the favored quarks are $\bar
{u}$ and $d$. The FFs of strange quarks and gluons  are left unmodified. Similar modifications were applied to obtain FFs into  $K^{+}$ and $K^{-}$,  but this time, the starting functions were the ones fragmenting $
u, \bar{u}, s, \bar{s}$ into the sum $K^{+}+K^{-}$.  Figure \ref{fig:NLO} shows that the  agreement between the NLO calculations that include the mKKP FFs and the measured pion cross section is remarkable (within 20\% above 1.5 GeV/c) even though the modifications applied can be considered as ``crude". Similar good agreement was obtained  for neutral pions at y=0 \cite{PHENIXpi0} and at y=3.8 \cite{STARppForward} at RHIC.
The agreement between the calculated and the measured kaon cross-sections is equally good. The difference between the mKKP and K parametrizations is driven by higher contributions from gluons fragmenting into pions. This difference has been identified as 
an indication that the gg and gq processes dominate the interactions at mid-rapidity \cite{PHENIXpi0}; these results indicate that such continues to be the case at high rapidity. The calculation that uses the K set underestimates the pion yields by a factor of $\sim 2$ at all values of $p_{T}$ while for positive kaons, the agreement is good at low momentum but deteriorates at higher momenta.

 An updated version of FFs that we refer to as the ``Albino, Kniehl and Kramer'' (AKK) set  has been extracted from more data made available recently  \cite{AKK}.
 It reproduces well the $p + \bar{p}$ distributions measured at mid-rapidity by the STAR collaboration \cite{STARppProton}. At high rapidity, the contribution from gluons fragmenting
into $p$ or $\bar{p}$ is dominant in this new set of FFs ($\geq 80\%$ for $p_{T} < 5$ GeV/c  \cite{Werner}), and the calculated cross sections for both  
particles consequently have nearly the same magnitude. We thus compare the measured cross sections for  $p$ and $\bar{p}$ to the NLO calculation using the
AKK FFs divided by 2 in the right-most panel of Fig. \ref{fig:NLO}. The calculation is close to the measured $p$ cross section but it is almost an order of magnitude higher than the measured $\bar{p}$ distribution. We conclude that the AKK FFs cannot be used to describe baryon yields at high rapidity because they fail to reproduce the measured abundance of $\bar{p}$ with respect to $p$. We have ruled out the use of the standard KKP FFs because they produce $p+\bar{p}$ cross sections that are  smaller by a factor of $\sim 10$ compared to the measurement (not shown). 
  
In summary,  unbiased invariant cross sections of identified charged particles as function of $p_{T}$ were measured at high rapidity in p+p collisions at $\sqrt{s}=\unit[200]{GeV}$. 
NLO pQCD calculations reproduce reasonably well the produced particle (pions and kaons) distributions  but $p$ and $\bar{p}$ cannot simultaneously be described well by any of the available FFs.  
These results may show a limitation of the factorized description of p+p cross sections, perhaps because it does not include the effects of baryon number transport that, as the data suggest, may extend to high $p_{T}$. These measurements 
bring additional insight into the hadron-hadron interaction and its description in the context of QCD; they are as well instrumental in constraining phenomenological descriptions of that system.

We thank Werner Vogelsang for providing us with the NLO pQCD calculations shown in this letter as well as many fruitful discussions during the preparation of this manuscript. 
This work was supported by 
the Office of Nuclear Physics of the U.S. Department of Energy, 
the Danish Natural Science Research Council, 
the Research Council of Norway, 
the Polish State Committee for Scientific Research (KBN) 
and the Romanian Ministry of Research.


\bibliography{apssamp}

\end{document}